\documentclass[aps,prl,twocolumn,groupedaddress]{revtex4-1}
\usepackage{bm}
\usepackage[dvipdfmx]{graphicx}
\begin{document}


\title{Quantum degenerate mixtures of alkali and alkaline-earth-like atoms}


\author{Hideaki Hara,$^1$ Yosuke Takasu,$^1$ Yoshifumi Yamaoka,$^1$ John M. Doyle,$^2$ Yoshiro Takahashi$^{1,3}$}
\affiliation{$^1$Department of Physics, Graduate School of Science, Kyoto University, Kyoto 606-8502, Japan\\
$^2$Harvard University, Department of Physics, Cambridge MA 02138\\
$^3$CREST, Japan Science and Technology Agency, Chiyoda-ku, Tokyo 102-0075, Japan
}


\date{\today}

\begin{abstract}
We realize simultaneous quantum degeneracy in mixtures consisting of the alkali and alkaline-earth-like atoms Li and Yb.
This is accomplished within an optical trap by sympathetic cooling of the fermionic isotope $^6$Li
with evaporatively cooled bosonic $^{174}$Yb and, separately, fermionic $^{173}$Yb.
Using cross-thermalization studies, we also measure the elastic s-wave scattering lengths of both Li-Yb combinations,
$|a_{^{6}\text{Li}-^{174}\text{Yb}}|=1.0 \pm 0.2$ nm and  $|a_{^{6}\text{Li}-^{173}\text{Yb}}|=0.9 \pm 0.2$ nm.
The equality of these lengths is found to be consistent with mass-scaling analysis.
The quantum degenerate mixtures of Li and Yb, as realized here, can be the basis for creation of ultracold molecules with electron spin degrees of freedom, studies of novel Efimov trimers, and impurity probes of superfluid systems.
\end{abstract}


\maketitle

Bose-Einstein condensation (BEC)\cite{BEC} and Fermi-degeneracy\cite{FD} in dilute atomic gases have provided deep insight into quantum many-body systems.
This includes studies of the crossover between a molecular BEC and the Bardeen-Cooper-Schriefer state\cite{crossover} and the strongly correlated state of Mott insulator in an optical lattice\cite{Mott} (for a full review see Ref.\cite{BlochRMP}).
Ultracold atomic gas mixtures of different elements offer many new intriguing possibilities 
such as dipolar physics with polar molecules\cite{polarmolecule}, study of heteronuclear Efimov resonance\cite{heteroEfimov}, and exploration of novel quantum states\cite{quantumphase}. 
To reach these goals, it is crucial to achieve simultaneous quantum degeneracy in an elementally mixed gas. 
While this has been achieved in systems with equivalent electronic structure
(one-electron atom combinations used to create dimers of elementally different alkali metal atoms), mixed structure gases have been notably absent.

Quantum degenerate systems consisting of both one-electron and two-electron atoms are of unique interest.
In particular, and in contrast to bi-alkalis, dimers made of alkali and alkaline-earth-like atoms not only have an electric dipole moment but also electronic spin degrees of freedom in the ground state\cite{RbYb_theory1,RbYb_theory2,PengZhang,Geetha}.
Such a spin-doublet molecule is important for the implementation of a proposed powerful spin-lattice quantum simulator\cite{latticespinmodel}.
Two ingredients are necessary to create high densities of such molecules. 
First is the availability of a magnetic Feshbach resonance.
Although not yet experimentally verified, these have recently been calculated and predicted to exist for an alkali and alkaline-earth system\cite{PhysRevLett.105.153201}.
Second is the production of high-phase-space dual species gas of the precursor atoms, which has remained an elusive goal.
We began initial work towards this goal by demonstrating a magneto-optical trapping (MOT) of our candidate atoms Li and Yb\cite{Okano, *[{Experimental works using RbYb are reported in }][{ ; }] YbRb,*YbRbPA}.

These particular elements are not only the right precursors for the creation of a spin-doublet molecule, but also have interesting properties due to the large inherent mass ratio of 29.
For example, recent theory predicts a novel collisional stability for weakly bound heteronuclear molecules formed in a two-species mixture with a large mass difference and light fermionic atoms\cite{stability}.
In addition, for a weakly bound three-body system with very asymmetric mass ratio such as LiYbYb, several consecutive Efimov states may exist\cite{efimov1973}.
Furthermore, mixtures with large mass difference nicely mimic an impurity problem in superfluids, where a heavy atom of Yb plays an role of the impurity and the light Li atom the superfluid\cite{impurity}.
Finally, novel quantum phases could be explored with mass-imbalanced mixture in a trap or an optical lattice\cite{FFLO, FFLO2}.

In this Letter, we report the first production of quantum degenerate Bose-Fermi and Fermi-Fermi mixtures of Yb and Li atoms.
This is realized in an optical trap using sympathetic cooling of a fermionic $^6$Li gas by an evaporatively cooled bosonic $^{174}$Yb and, separately, multi-component fermionic $^{173}$Yb gas. 
Highly efficient cooling leads to a low temperature of the $^6$Li gas to as low as about 0.1 of the Fermi temperature $T_F$.
The coolant Yb cloud also provides excellent thermometry for this very cold Fermi gas.
In-trap cross-thermalization between the two species is used to measure the absolute value of s-wave scattering lengths
$|a_{^{6}\text{Li}-^{174}\text{Yb}}|=1.0 \pm 0.2$ nm and $|a_{^{6}\text{Li}-^{173}\text{Yb}}|=0.9 \pm 0.2$ nm.
The equality of these values is consistent with our mass-scaling analysis of scattering lengths, as discussed later in this paper.

The basic scheme of our experiment is to sympathetically cool fermionic $^6$Li with Yb.
Figure 1 shows our experimental setup and timing chart.
\begin{figure}
\includegraphics[width=7cm,height=10cm]{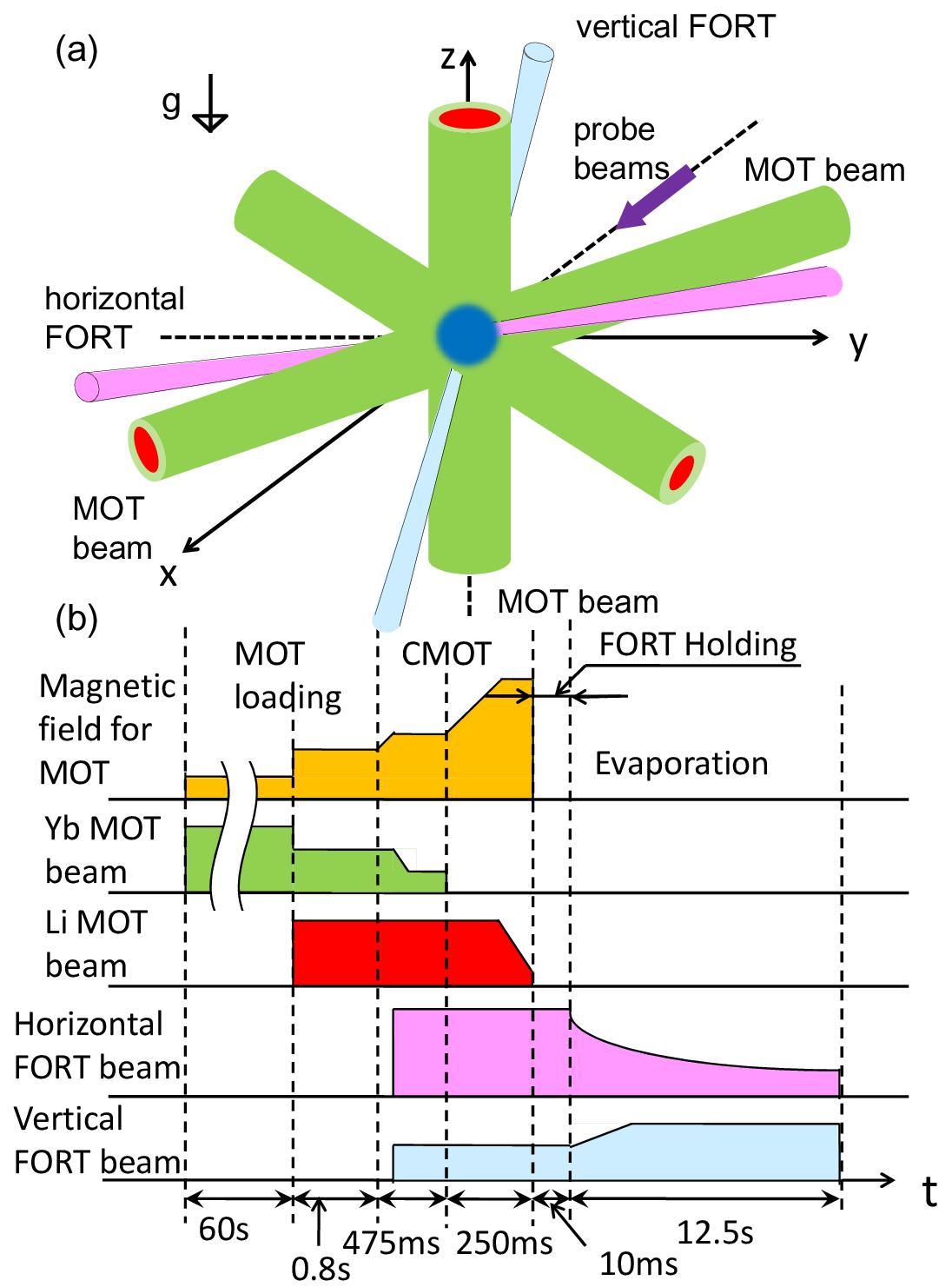}%
\caption{\label{timechart}Experimental setup (a) and procedure (b)
to trap atoms in MOT and FORT, and cool the mixture down to the quantum degenerate regime. Typical time duration is also shown.
CMOT represents a compressed MOT.}
\end{figure}
The experiment begins with a simultaneous MOT of $^6$Li and Yb.
The apparatus and scheme for the MOT is an extension of our previous setup described in Ref.\cite{Okano}.  
We use a dual atomic oven, which contains both Li and Yb.
For $^6$Li, we use the $^2$S$_{1/2}$-$^2$P$_{3/2}$ (D$_2$) transition with
the wavelength of 671 nm for Zeeman slowing, MOT, and probing.
For Yb, the $^1$S$_0$-$^1$P$_1$ transition ($399$ nm) is used for Zeeman slowing and probing, 
while the $^1$S$_0$-$^3$P$_1$ intercombination transition ($556$ nm) is used for the MOT, in order to obtain a very cold sample of Yb.
The typical numbers of atoms in our MOT are $N_{\text{Li}}=3.7 \times 10^8$ for $^6$Li and $N_{\text{Yb}}=9.0 \times 10^6$ for $^{174}$Yb.

To obtain a dense cloud for evaporative cooling, we transfer the atoms from the MOT into a far-off-resonant optical trap (FORT) (see Fig. 1(b)).
The horizontally propagating laser beam of the FORT is generated by a fiber laser with the wavelength of 1070 nm and power tunable by an acousto-optic modulator.
The maximum FORT power is 36 W with beam waist $w_0$ of 24 $\mu$m, estimated by parametric resonance measurements.
This results in a potential depth of 1.3 mK for Yb and 2.5 mK for Li.
This difference in the trap depth is quite favorable to achieve a low temperature of Li atoms through sympathetic cooling with Yb atoms.
The Yb atoms evaporate preferentially, in turn cooling the deeply trapped Li.
To achieve tighter trap confinement, another FORT beam (1083 nm, 25W, and beam waist of 75 $\mu$m) is applied along the near vertical direction (see Fig. 1(a)).

Since $^6$Li atoms are difficult to be evaporatively cooled by themselves, due to their small elastic scattering cross section at low magnetic field, collisions between Li and Yb dominate the cooling of Li atoms in our system.
We take special care to load an appropriate number of $^6$Li atoms into the trap; too many $^6$Li atoms will cause a severe loss of Yb atoms during sympathetic evaporative cooling.
We typically load $^6$Li atoms in a MOT for only 0.8 s after loading Yb for 60 s, as shown in Fig. 1(b).
The compressed MOT phase with $10$ ms holding time follows, during which time both Li and Yb atoms are transferred into the FORT.
The Li atoms are optically pumped into the $F=1/2$ state to avoid inelastic losses in the FORT.
We perform evaporative cooling by gradually decreasing the trap depth of the horizontal FORT beam.
The time duration of evaporative cooling was optimized for each Yb isotope.
For the $^{174}\text{Yb}$ and $^{6}\text{Li}$ mixture, we perform evaporative cooling by lowering the horizontal FORT beam power from 36 W to 0.2 W over 12.5 s.
\begin{figure}
\includegraphics[width=8cm,keepaspectratio]{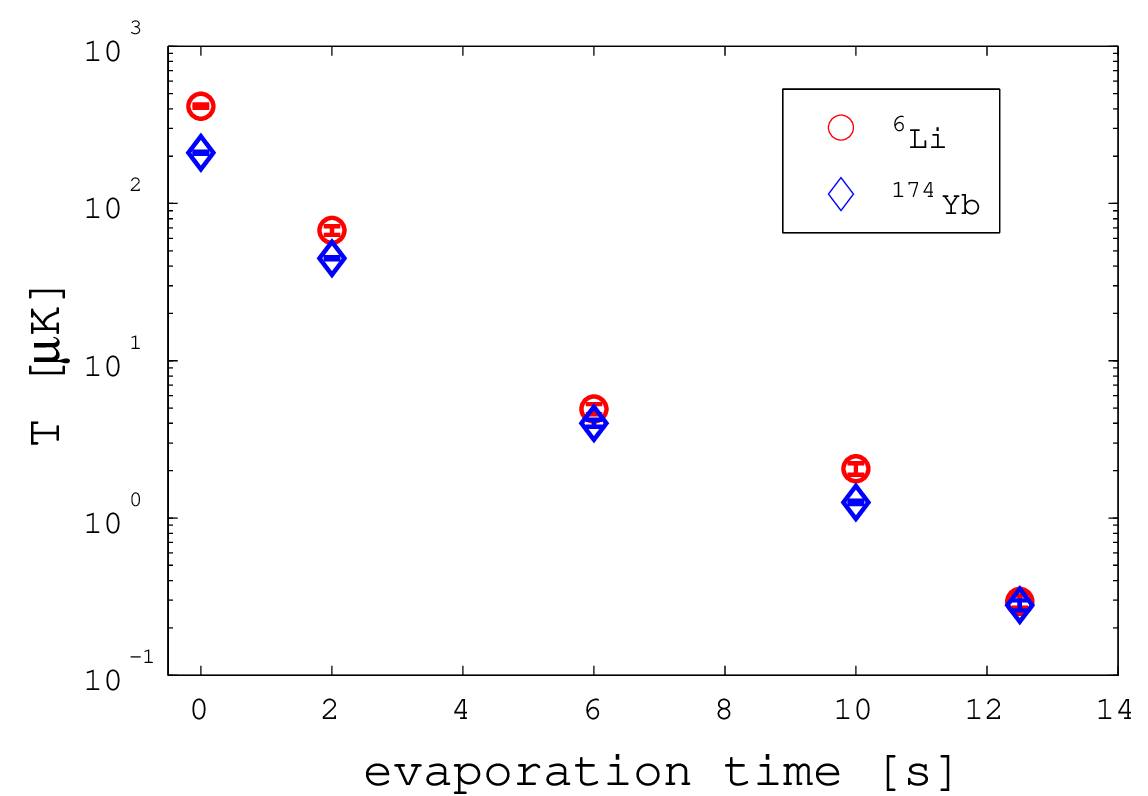}%
\caption{\label{evaporation}Temperature evolution of $^6$Li (circle) and $^{174}$Yb (diamond) during the evaporative cooling.
A thermal contact between $^6$Li and $^{174}$Yb is obtained during the evaporative cooling.}
\end{figure}
The temperature locking between Li and Yb, shown in  Fig. 2, in combination with the very high trap depths for Li, clearly indicates good thermal contact between Li and Yb.

\begin{figure}[b]
\includegraphics[width=9cm,keepaspectratio]{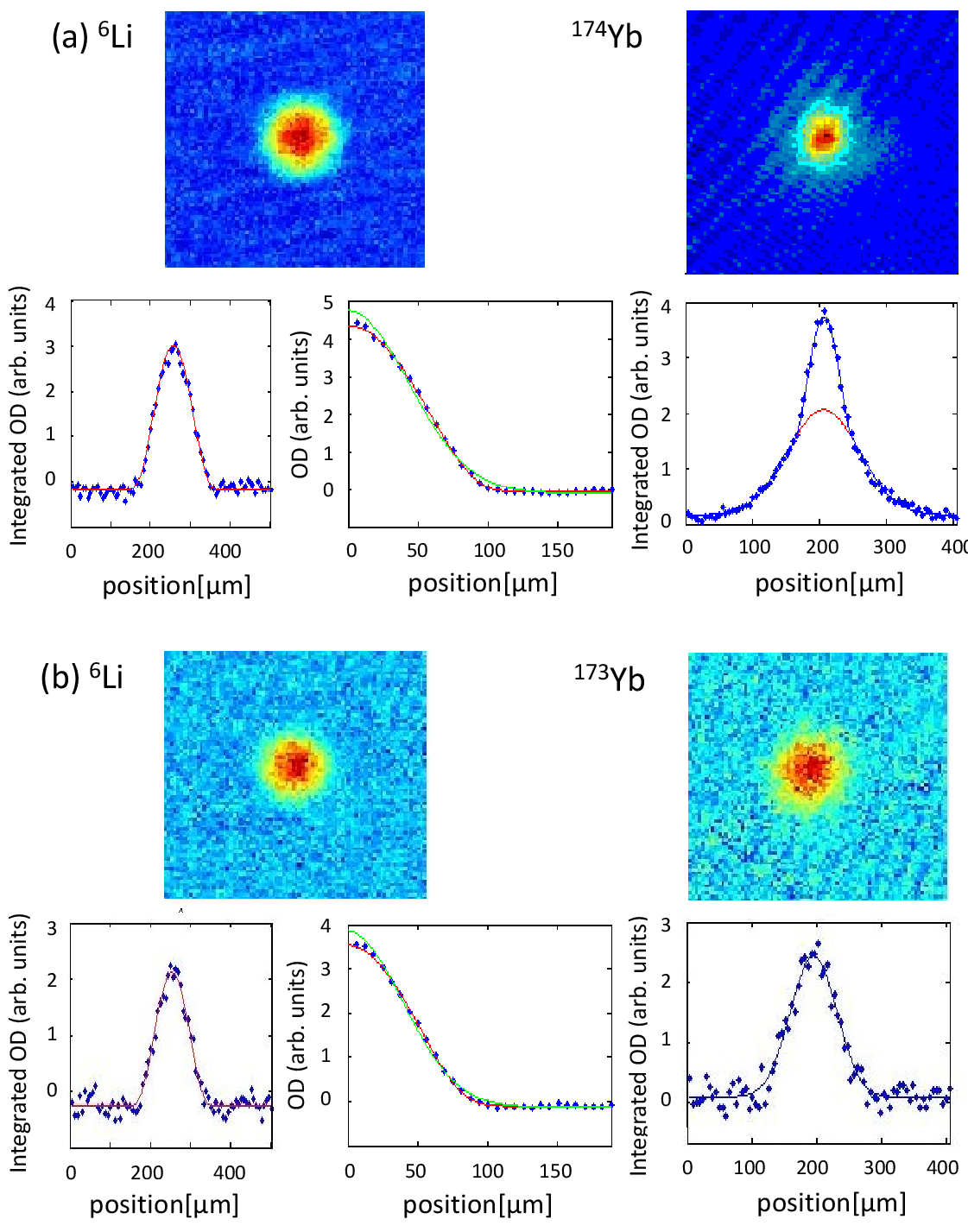}%
\caption{\label{imaging}Time of Flight absorption images of the quantum degenerate Bose-Fermi mixture of $^6$Li and $^{174}$Yb ((a) left and right), and Fermi-Fermi mixture of $^6$Li and $^{173}$Yb ((b) left and right).
(a, left)The optical column density(upper) and its projection(lower left) are shown as well as the azimuthally averaged distribution(lower right) for $^6$Li in the Bose-Fermi mixture.
Expansion time is 1 ms.
These data are averaged over 10 measurements.
The temperature of $0.1T_F$ is determined from the Thomas-Fermi fit(red line). 
The observed distribution clearly deviates from the classical Gaussian shape, indicated by the green line.
(a, right)The optical column density(upper) and its projection(lower) are shown for $^{174}$Yb in the Bose-Fermi mixture.
Expansion time is 16 ms.
These data are averaged over 10 measurements.
The bimodal distribution is clearly observed.
(b, left)Similar plots as in (a, left) but for $^6$Li in the Fermi-Fermi mixture.
Expansion times is 1 ms.
The data are averaged over 5 measurements.
The temperature of $0.1T_F$ is determined from the Thomas-Fermi fit(red line). 
The observed distribution clearly deviates from the classical Gaussian shape, indicated by the green line.
(b, right)Similar plots as in (a, right) but for $^{173}$Yb in the Fermi-Fermi mixture.
Expansion time is 12 ms.
The data are averaged over 8 measurements.
OD means the optical density.
}
\end{figure}
The Time-of-Flight (TOF) images in the final phase of the experiment, as shown in Fig. 3, indicate simultaneous quantum degeneracy in the Li and Yb gases.
Figure 3(a) shows the results for the Bose-Fermi mixture.
The total $^6$Li atom number $N_{\text{Li}}$ is $2.5 \times 10^4$, which is equally distributed between the $|+1/2>$ and $|-1/2>$ states. 
The Fermi temperature is estimated as $T_{F}= 3.8 \pm 0.6$ $\mu$K, extracted using the mean trap frequency of $\omega_{\text{Li}} = 2 \pi \times 1.8$ kHz.
Fitting the TOF image to a Fermi-Dirac distribution results in $T_{\text{Li}}=290 \pm 30$ nK, corresponding to $T/T_{F}=0.08\pm0.01$, deep into the Fermi degenerate regime.
This value is consistent with $T/T_{F}=0.08\pm0.02 $ extracted from the fugacity.
The temperature of the $^{174}$Yb is 280 $\pm$ 20 nK, below the calculated BEC transition temperature of $T_C=510$ nK. 
The bimodal distribution of $^{174}$Yb shown in Fig. 3(a, right) indicates successful simultaneous formation of $^{174}$Yb BEC.
The number of atoms in the condensate is $N_{\text{BEC,Yb}} = 1.5 \times 10^4$.
The relatively large non-condensed fraction, compared to a pure sample, seems to be due to the presence of the $^6$Li gas, 
which is cooled by the $^{174}$Yb.
This phenomenon is open to systematic study in a future experiment.
We show in Fig. 3(b) the results for the Fermi-Fermi mixture.
The hyperfine spin of $5/2$ for $^{173}$Yb results in a $6$ spin component gas of $^{173}$Yb.
Fitting the TOF images to Fermi-Dirac distributions gives $T_{\text{Li}}=220 \pm 40$ nK and $T/T_{F}=0.07 \pm 0.02$ for $^6$Li and $T_{\text{Yb}}=170 \pm 10$ nK and $T/T_{F}=0.52 \pm 0.12$ for $^{173}$Yb.
The equality of the temperatures between the Yb and Li atoms within the experimental error indicates that the Yb cloud provides excellent thermometry for this very cold Fermi gas of $^6$Li.
It is also noted that both $^{173}$Yb and  $^6$Li have spin degrees of freedom (sixfold and twofold, respectively).
Thus this system offers another unique quantum degenerate fermionic system with novel spin symmetry\cite{taie}.

In addition to mixed quantum gases, we also produce single-species quantum degenerate gases.
A $^{174}$Yb BEC with $N_{\text{BEC,Yb}}=2.0\times 10^4$ and a $^{173}$Yb Fermi degenerate gas with $T/T_{F}=0.39 \pm 0.09$ are obtained in the absence of $^6$Li.
These pure Yb quantum gases are notable as they have a lower photon scattering rate compared with the Yb gases so far realized using a 532 nm FORT.
We also obtain the single-species Fermi degenerate gas of $^6$Li by evaporating away all the coolant $^{174}$Yb atoms, after degeneracy.
$N_{\text{Li}}=7.7 \times 10^4$ and $T_{\text{Li}}=1.1 \pm 0.1$ $\mu$K, corresponding to $T/T_{F}=0.20 \pm 0.04$, which is consistent with $T/T_{F}=0.21 \pm 0.02$ extracted from the fugacity.

We observe cross-thermalization between $^6$Li and $^{174}$Yb, as well as between $^6$Li and $^{173}$Yb.
We do this by measuring the temporal evolution of the temperatures of the Li and Yb atoms in a horizontal FORT beam with the trap depth kept constant.  
While the temperature of $^6$Li atoms is almost constant in the absence of Yb atoms, 
the $^6$Li temperature goes down from 500 $\mu$K to 300 $\mu$K with a time constant of about 0.5 s in the presence of $^{174}$Yb atoms.
Similarly, the $^6$Li gas is cooled from 400 $\mu$K to 250 $\mu$K with a time constant of about 0.8 s by $^{173}$Yb atoms.
This is the evidence of cross-thermalization between Li and Yb.
From the measured Li thermalization time constants, we determine the interspecies cross section $\sigma_{\text{Li-Yb}}$\cite{thermalization}.
Since the p-wave centrifugal barrier is 2.8 mK, only the s-wave scattering occurs at the temperatures of our experiment.
From the formula of $\sigma_{\text{Li-Yb}}=4 \pi {a_{\text{Li-Yb}}}^{2}$, the absolute values of the  s-wave scattering lengths between $^6$Li and $^{174}$Yb and between $^6$Li and $^{173}$Yb are measured to be $\left|a_{\text{$^6$Li-$^{174}$Yb}}\right| = 1.0 \pm 0.2$ nm and $\left|a_{\text{$^6$Li-$^{173}$Yb}}\right| = 0.9 \pm 0.2$ nm, respectively.
The result for $\left|a_{\text{$^6$Li-$^{174}$Yb}}\right|$ is consistent with a recently reported value\cite{gupta}.

The quite small isotopic variation in the measured scattering lengths for $^6$Li -$^{174}$Yb and $^6$Li -$^{173}$Yb is in contrast with our previous observation of a large isotopic variation of scattering lengths for Yb-Yb\cite{kitagawa}. 
This difference can be explained by the same physical mass-scaling effect but with a drastically different mass ratio between
the collision partners (1:1 for Yb-Yb, versus about 29:1 for Yb-Li).
The $s$-wave scattering length is given by the following formula\cite{GF1,Flambaum,Boisseau},
\begin{equation}
a = \bar{a} \left[ 1-\tan\left(\Phi-\frac{\pi}{8}\right)\right].
\label{massscaling}
\end{equation}
Here $\bar{a}=2^{-3/2}\frac{\Gamma(3/4)}{\Gamma(5/4)}\left(2\mu C_6/\hbar^2\right)^\frac{1}{4}$ is a characteristic length associated with the van der Waals potential,
where $\Gamma$ is the gamma-function, $\mu $ is the reduced mass, and $\hbar $ is the Planck constant divided by $2\pi $.
The calculated value of $C_6$\cite{PengZhang, Geetha} gives $\bar{a}= 2$ nm.
The semiclassical phase $\Phi$ is defined by
$\Phi = \frac{\sqrt{2\mu}}{\hbar}\int_{r_0}^{\infty} \sqrt{-V(r)}dr$,
where $r_0$ is the inner classical turning point of $V(r)$ at zero energy.  
The number of bound states $N$ in the potential is~\cite{Flambaum}
$ N=\left [ \frac{\Phi}{\pi} -\frac{5}{8} \right ] +1$, 
where the brackets mean the integer part. 
Recent {\it ab initio} calculations\cite{PengZhang, Geetha} of the LiYb potential shows that the phase is about 80 rad. 
Due to the quite small difference in the reduced masses for $^6$Li -$^{174}$Yb and $^6$Li -$^{173}$Yb, the difference in the phase 
 is calculated to be less than 0.01 rad, which results in almost the same scattering lengths for both cases.

In conclusion, we produce and study quantum degenerate Bose-Fermi mixtures of $^6$Li and $^{174}$Yb and Fermi-Fermi mixtures of $^6$Li and $^{173}$Yb.
The fermionic $^6$Li gas is sympathetically cooled by evaporatively cooled Yb.
The $^6$Li gas reaches the deep Fermi degenerate regime with a temperature of only about $0.1T_F$, while coexisting with the $^{174}$Yb BEC or $^{173}$Yb Fermi degenerate gas.
The absolute values of s-wave scattering lengths $|a_{^{6}\text{Li}-^{174}\text{Yb}}|=1.0 \pm 0.2$ nm and  $|a_{^{6}\text{Li}-^{173}\text{Yb}}|=0.9 \pm 0.2$ nm are obtained and analyzed using the mass-scaling analysis.
The realized quantum degenerate mixtures of Li and Yb atoms open the door to the study of unexplored quantum few-body and many-body systems. 

We acknowledge M. Okano, M. Muramatsu, S. Uetake for their experimental assistance.
We thank H. Sadeghpour, P. Zhang, A. Dalgarno, G. Gopakumar, and M. Abe for helpful discussions.
This work was supported by the Grant-in-Aid for Scientific Research of JSPS (No. 18204035, 21102005C01 (Quantum Cybernetics), 21104513A03 (DYCE), 22684022), GCOE Program "The Next Generation of Physics, Spun from Universality and Emergence" from MEXT of Japan, FIRST, and Matsuo Foundation.

\begin{acknowledgments}

\end{acknowledgments}


\providecommand{\noopsort}[1]{}\providecommand{\singleletter}[1]{#1}%

\end{document}